# Title: Radio Jets Clearing the Way Through a Galaxy: Watching Feedback in Action

**Authors:** Raffaella Morganti[1,2]*, Judit Fogasy[3], Zsolt Paragi[4], Tom Oosterloo[1,2], Monica Orienti[5]

**Affiliations:**

[1] ASTRON, Netherlands Institute for Radio Astronomy, Postbus 2, 7990 AA, Dwingeloo, The Netherlands

[2] Kapteyn Astronomical Institute, University of Groningen, P.O. Box 800, 9700 AV Groningen, The Netherlands

[3] Eötvös Loránd University, Egyetem tér 1-3., 1053 Budapest, Hungary

[4] Joint Institute for VLBI in Europe, Postbus 2, 7990 AA Dwingeloo, Netherlands

[5] INAF - Istituto di Radioastronomia, via Gobetti 101, I-40129, Bologna, Italy

* Corresponding author: R. Morganti (morganti@astron.nl)

**Abstract**: The energy released by an active galactic nucleus (AGN) has a strong impact on the surrounding interstellar medium (ISM). This feedback is considered to be the regulating factor for the growth of the central massive black hole, and for the rate of star formation in a galaxy. We have located, using Very Long Baseline Interferometry (VLBI), the fast outflow of neutral hydrogen in the young, restarted radio loud AGN 4C12.50. The outflow is located 100 pc from the nucleus where the radio jet interacts with the ISM, as well as around the associated radio lobe. These observations show that the radio plasma drives the outflow and removes gas from the central regions, and that jet driven outflows can play a relevant role in feedback mechanisms.

**Main Text:** The important role for galaxy evolution of energetic feedback effects due to activity in the galaxy's nucleus has been put, in recent years, on more solid grounds thanks to the discovery of massive gas outflows in a growing number of galaxies with AGN. Many questions are still open, however, in particular regarding the nature of the driving mechanism of these outflows (e.g. *1, 2*). Answering these questions is important because it has implications for how ubiquitous the AGN related feedback is, whether it is connected to specific phases in the life of an AGN and whether it is a recurrent phenomenon. One of the best ways to answer these questions is to identify the location of the outflowing gas and to image its distribution and kinematics. So far, this has not been achieved given the parsec scale spatial resolution required to resolve the gas outflow.

The kinetic push of radio jets is often considered a possible mechanism for driving a gas outflow because of the high efficiency with which jets can transfer energy to the ISM. However, their narrow opening angle is often used as an argument that their impact cannot be very high because a narrow jet would affect only a small part of the ISM.Yet, recent numerical simulations (*3, 4*) have shown that a radio jet, especially when the radio source is in an initial phase and surrounded by a porous, clumpy medium, may be able to efficiently clear up the gas in which it is enshrouded. This is because a large cocoon of disturbed and outflowing gas is created around the jet by the interaction of the radio plasma, thus affecting a much larger region of the galaxy.

We have imaged, on parsec scales, the distribution and kinematics of the fast outflowing component of the neutral hydrogen in 4C12.50, one of the best-known Ultraluminous Infrared Galaxies (ULIRG) that hosts a young – recently restarted – radio-loud AGN. Galaxies like 4C12.50 are particularly relevant because they are considered to be the link between ULIRG and AGN (*5*); hence, they represent a particularly interesting phase in the evolution of a galaxy. The intriguing and fascinating characteristic of 4C12.50 is the presence of fast outflows not only of ionized gas (*6, 7*), but also of cold gas, in particular atomic hydrogen (HI) and molecular gas (CO). An HI outflow of ~1000 km s$^{-1}$ has been previously detected (*8*) that was later found (9) to have a markedly similar counterpart of molecular gas (CO(1-0) and (3-2)), illustrated in Fig. 1. These gaseous components have been interpreted as being part of the same fast outflow of cold gas.

In 4C12.50, all the possible mechanisms that could drive fast gas outflows are potentially present (*10*): starburst wind (connected to the relatively young stellar population), radiative AGN wind-driven outflows (connected to the bright optical AGN with high-ionization gas), and a powerful radio source. The presence of a particularly strong interaction between the radio jet and the ISM has been suggested by the presence of a hot spot with a very high fractional polarization (60%) at the end of the radio jet (*11*). Although the starburst wind has been ruled out as driving the outflow (*10*), the other possibilities remain. To locate the outflowing gas we have performed observations with milliarcsecond (mas) angular resolution, using Very Long Baseline Interferometry (VLBI).

The observations were performed using a global VLBI network, including the Very Long Baseline Array (VLBA), one antenna of the Very Large Array (VLA) and three European VLBI Network (EVN) telescopes (Effelsberg, Westerbork and Onsala). The central frequency, corresponding to the frequency of the redshifted HI, was set at 1266 MHz. These observations expand on the results obtained previously (*12*) which were limited by the width of the observing band. We have used a broader observing band (16 MHz, covering gas with kinematics deviating more than 1500 km/s from the systemic velocity) and a longer integration time (14h). The angular resolution ranges from 12x4 m.a.s. (PA 12$^o$) for the naturally weighted images and cubes, to 8x5 m.a.s. (PA 2$^o$) for the full-resolution images. They correspond, at the redshift of 4C12.50 z=0.1217, to 32x11 pc and 21x13 pc respectively. The noise level obtained is 0.65 mJy/b/ch and 0.90 mJy/b/ch respectively for the different weightings, after Hanning smoothing. A continuum image was also obtained using the line-free channels (Fig. 2).

We detect the two HI absorption features previously seen in lower-resolution observations (*8*) and imaged them at parsec scale resolution (Fig. 1). The outflowing component (~1000 km s$^{-1}$ blueshifted from the systemic velocity) is seen at the end of the distorted southern jet. This HI component appears to consist of a compact cloud (unresolved even by our high-resolution images) and a diffuse tail-like structure. The compact cloud is seen – in projection – to be co-spatial with the hot-spot observed in radio continuum (*11*). The faint and diffuse component extends at least ~50 pc around and in front of the southern lobe (Fig. 1). This can be considered a lower limit to the actual extent due to the sensitivity limitations of the observations. Indeed, the profile from the naturally weighted image integrated over the southern lobe shows that we recover most of the HI absorption detected with the Westerbork Synthesis Radio Telescope (WSRT, Fig. 3). The deeper HI absorption feature at the systemic velocity detected earlier (12) is confirmed by our observations to correspond to a cloud north of the nucleus (Fig. 1).

We have estimated the column density and the HI mass associated with the two clouds. In the southern part, the unresolved cloud has a column density of $N_{HI} = 4.6 \times 10^{21}$ cm$^{-2}$ (assuming the temperature $T_{spin}$=100 K). The derived mass of the cloud is $M_{HI} = 600$ $M_\odot$; if we include the extended part, it reaches $M_{HI} = 1.6 \times 10^4$ $M_\odot$. The assumed value of $T_{spin}$ likely represents a lower limit because the actual spin temperature, under the extreme conditions close to the AGN, may be at least a factor of 10 larger (1*6, 13*). This implies that also the mass of the cloud, as well, could be much higher. A recent study (*9*) suggests that the outflow carries an estimated cold $H_2$ mass of at least $4.2 \times 10^3$ $M_\odot$.

The larger HI cloud located in the northern part of the radio source is known to have an extremely high optical depth ($\tau = 0.6$) *(12)*. This is confirmed by our observations. The column density that we derived reaches $N_{HI} = 4.4 \times 10^{22}$ cm$^{-2}$ (again assuming the conservative value of $T_{spin}$=100 K) and the corresponding mass of the cloud is $1.4 \times 10^5$ $M_\odot$, which is slightly higher than previously derived (*12*). Indeed, for this cloud the broader band used in our observations has helped in recovering part of the HI that was missed before (in particular a broad, blueshifted wing). The cloud appears clearly resolved and we derived a size of 50 x 80 pc. This represents a lower limit because the cloud could be more extended, outside the region covered by the radio continuum where the HI cannot be traced in absorption.

For a number of radio sources, it has been proposed that the radio jet is responsible for the fast gas outflows. For 4C12.50 we could pinpoint the location of the outflow gas and recover the distribution of the cold gas associated with the most blueshifted HI absorption in relation to the radio jet. The extreme kinematics, together with the location (in projection) of the HI, suggests that we are indeed witnessing gas being expelled from the galaxy at a (projected) speed of 1000 km/s as a result of the interaction between the radio jet and a dense cloud in the ISM. The presence of such a strong interaction between the radio jet and the ISM had already been proposed for 4C12.50 following the study of the warm, ionized gas (*6, 7*). The emission lines show a complex structure with prominent blueshifted wings that have been interpreted as gas disturbed by the interaction with the radio plasma. However, the high spatial resolution of our data allowed us to study this in more detail.

The marked similarity of the blueshifted HI and CO absorption profiles points to a similarity in the distribution and kinematics of the gas in these two phases. In the context of jet-ISM interactions, this would suggest that relatively dense clouds are present in the medium that do not get fully destroyed by such an interaction. The average densities derived for the compact clouds we observed range between 150 cm$^{-3}$ and 300 cm$^{-3}$. These values are similar to those for the clouds in numerical simulations where a jet drives the lateral expansion of the medium around the jet which, in turn, markedly affects the ISM of the host galaxy (*3,4*). Thus, in 4C12.50, we may be witnessing this process in action.

The cold gas in the compact cloud may represent the core of an originally larger cloud that is encountered by the jet. The extended, diffuse trail observed against and around the radio lobe would instead represent the expanding medium dispersed after the interaction. Because of the similarity between the HI and CO blueshifted absorption profiles, we suggest that the two phases of the gas come from the same structure, with the HI likely an intermediate phase in the cooling process and the CO ($H_2$) the final stage. This gas must have undergone rapid cooling to be observed both in its HI and molecular phase. By following previous models (*14*), we found the cooling time for the southern cloud in 4C12.50, using the densities and velocities from our data, to be on the order of $10^4$ yr. Such a time scale is comparable to the age of the radio source, suggesting that the gas has indeed time to cool during the process of being expelled from the central regions of the galaxy.

An interaction between the radio jet and a clumpy medium surrounding the jet was also speculated to explain the region of extremely high polarization, ranging up to 60% (*11*). The colocation of this region with extremely high fractional polarization with the HI blueshifted cloud supports the suggestion that this feature is a termination shock or ''working surface'' where the jet encounters the interstellar medium and rapidly decelerates. The high fractional polarization would result from the compression of a gas cloud and associated amplification of the magnetic field at that location.

The amount of energy in the outflow, as a fraction of the accretion energy of the AGN, is an important parameter for judging the relevance of the outflow. By combining the HI and optical results, it was found that the total observed outflow is between 16 and 29 $M_\odot$/yr, which would account for 0.2-0.3 per cent of the available accretion energy (*7*). This contrasts with the requirements of most quasar feedback models (5–10 per cent of the Eddington luminosity) but is similar to what is required in the "two phase" model proposed by (*1*). The morphology of the faint, diffuse gas seen in absorption in 4C12.50 is similar to that predicted by this model and suggests that we may indeed be witnessing the stretching and deformation of the ISM clouds required by (*1*) in order to increase the effective cross-section of the interaction.

The detection of an extended (total size ~150 kpc), low surface brightness radio structure around 4C12.50 (*15*), likely the signature of a previous cycle of activity, suggests that the feedback of the radio jets on the ISM may be a repeating effect in the life of this galaxy. Indeed, most of the other radio sources known to have fast outflows are – like 4C12.50 - young or restarted radio sources. This would underline the role of radio jets in feedback mechanisms because it implies that this type of feedback can act more than once in the life of a galaxy, and therefore is even more relevant in its evolution. It could be that the stronger impact of the radio plasma occurs in the initial/young phase of the radio jet, where the young radio source clears up the central regions (see e.g. the case of PKS1549-79)(*16*), and that at a later stage, the radio-loud activity corresponding with the 10-100 kpc size radio jet, provides a "maintenance" mode, with energy deposited at larger radii in the surrounding hot halo of the galaxy, providing a mechanism to prevent the gas in these halos from cooling (*17*).

**References and Notes:**
1. Hopkins P.F., Elvis M., 2010, *Mon. Not. R. Astron. Soc.*, **401**, 7

**Acknowledgments:** The European VLBI Network is a joint facility of European, Chinese, South African and other radio astronomy institutes funded by their national research councils. The National Radio Astronomy Observatory is a facility of the National Science Foundation operated under cooperative agreement by Associated Universities, Inc. JF acknowledges the ASTRON/JIVE Summer Student Programme during which she carried out this project. Z.P. acknowledges partial support from the Hungarian Scientific Research Fund, grant number OTKA K104539.

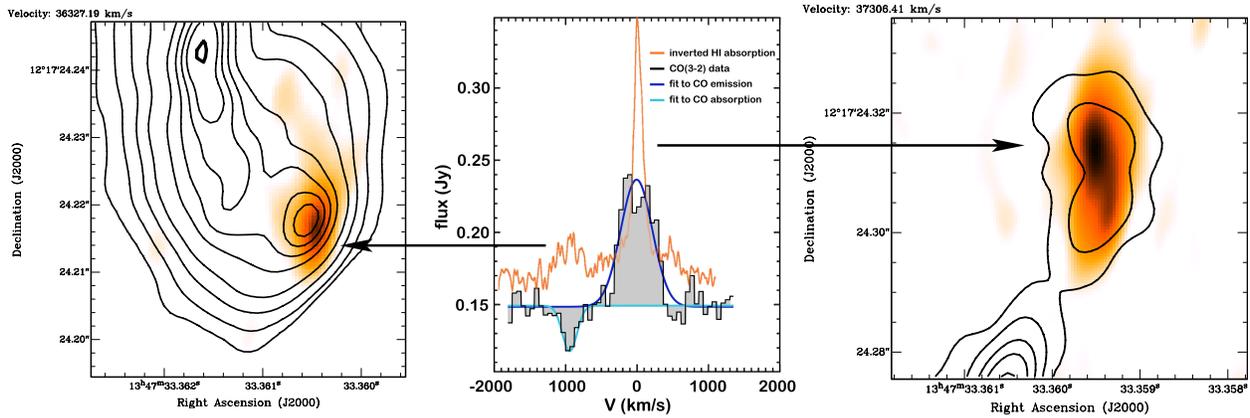

**Fig. 1** – The distribution of the HI in two velocity channels showing the location of the two clouds of HI detected in absorption (orange-white) superimposed to the continuum of 4C12.50 (contours). The integrated HI absorption profile is shown in the middle superimposed to the CO profile (taken from (*9*) with the HI from (*8*) and inverted for comparison purpose). The northern HI cloud is the one at the systemic velocity (*11*) whereas the southern one is blueshifted of ~1000 km/s compared to the systemic velocity. The location of the southern unresolved cloud producing the blueshifted HI absorption is co-spatial (in projection) with the bright radio hot spot. A diffuse HI component is also observed. Credit: Dasyra & Combes 2012, A&A 541, L7, reproduced with permission © ESO

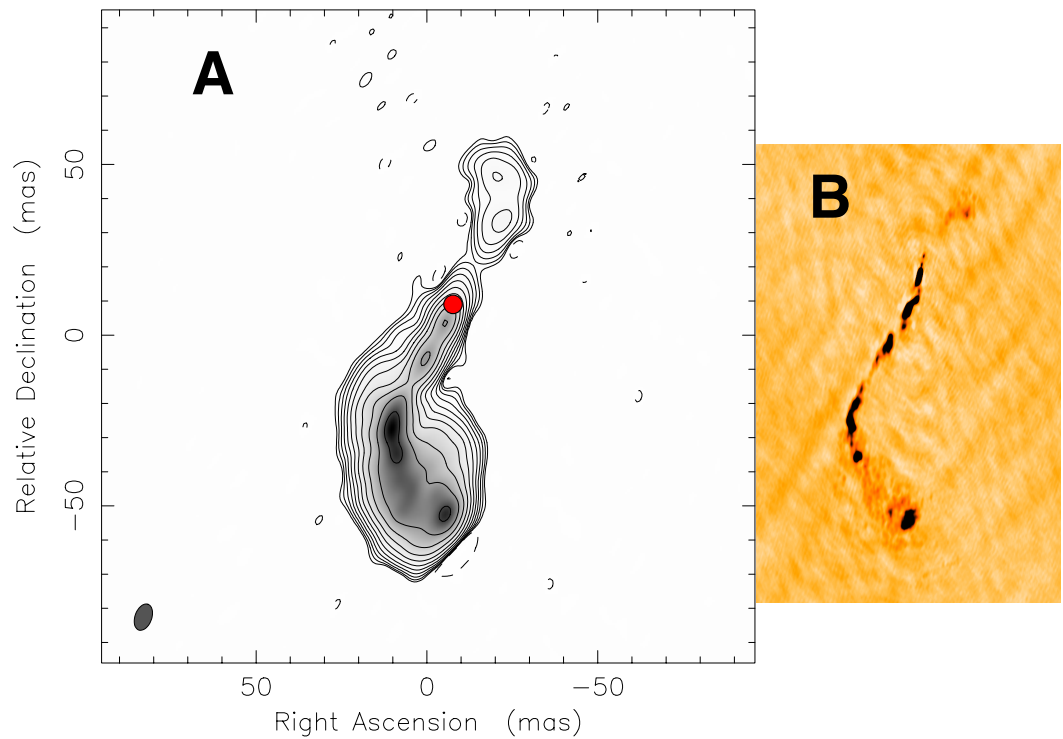

**Fig. 2**. Radio continuum images of 4C12.50 illustrating the morphology of the radio jet. (A) The continuum image derived from our VLBI data using the line-free channels. The red dot indicates the location of the radio core. The contour levels are 0.0003x (-1, 1, 2, 4, 8, 16, 32, 64, 128, 256, 512, 1024) Jy/beam with peak brightness 0.417 Jy/beam. (B) Continuum image taken from a previous, higher spatial resolution, study (*12*). The bent structure of the jet and the terminal hot spot in the southern lobe are clearly seen. This hot spot is characterized by an extremely high (~60%) fractional polarization (*12*). The two images are displayed on approximately the same scale. [Reproduced by permission of the American Astronomical Society.]

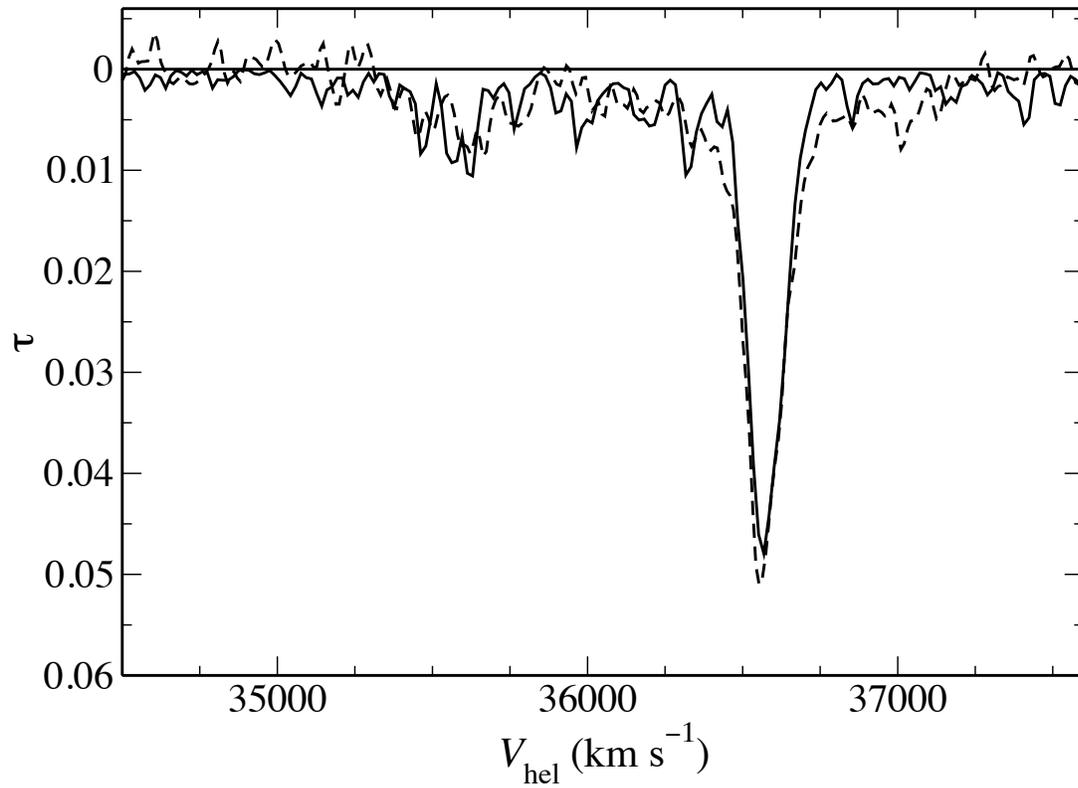

**Figure 3 -** Comparison of the HI absorption profiles from low and high-resolution observations. The dashed profile represents the HI absorption from the WSRT observations (*8*) and is shown superimposed on the integrated profile from the VLBI observations presented here (in black).